\documentstyle[doublespacing,referee]{mn}

\input epsf

\def\apj{ApJ}

\def\aj{AJ}
\def\aap{A\&\hskip-1pt A}

\def\mnras{MNRAS}

\def\lesssim{\mathrel{\hbox{\rlap{\hbox{\lower4pt\hbox{$\sim$}}}\hbox{$<$}}}}
\def\gtrsim{\mathrel{\hbox{\rlap{\hbox{\lower4pt\hbox{$\sim$}}}\hbox{$>$}}}}
\newcommand{\deltavec}{\mbox{\boldmath $\delta$}}
\newcommand{\thetavec}{\mbox{\boldmath $\theta$}}
\newcommand{\uvec}{\mbox{\boldmath $u$}}
\newcommand{\rvec}{\mbox{\boldmath $r$}}
\newcommand{\xvec}{\mbox{\boldmath $x$}}
\newcommand{\yvec}{\mbox{\boldmath $y$}}

\title[Planet Astrometric Microlensing]
	{Properties of Planet-induced Deviations in the
         Astrometric Microlensing Centroid Shift Trajectory}

\author[Han \& Lee]
{
Cheongho Han$^\dagger$ \& Chunguk Lee$^\ddagger$\thanks{e-mails: 
cheongho@astroph.chungbuk.ac.kr (CH); leecu@astro.chungbuk.ac.kr (LC)}\\
${}^\dagger$Department of Physics, Chungbuk National University, 
Chongju 361-763, Korea\\
${}^\ddagger$Department of Astronomy \& Space Science, 
Chungbuk National University, Chongju 361-763, Korea
}

\begin{document}
\maketitle
\label{firstpage}
\vspace{-\abovedisplayskip}
\begin{abstract}
An extra-solar planet can be detected by microlensing because it can 
distort the smooth lensing light curve created by the primary lens.  
As a new method to search for and characterize extra-solar planets, 
Safizadeh, Dalal \& Griest proposed to detect the planet-induced 
distortions in the trajectory of the microlensed source star's centroid 
motion (astrometric curve), which is observable by using the next 
generation of high-precision interferometers. In this paper, we 
investigate the properties of the planet-induced deviations in the 
astrometric curves (excess centroid shifts $\Delta\deltavec$) and the 
correlations of $\Delta\deltavec$ with the photometric deviations. 
For this, we construct vector field maps of $\Delta\deltavec$, which 
represent the difference of the centroid shift from that expected in the 
absence of the planet as a function of source positions.  From this 
investigation, we find that significant astrometric deviations occur 
not only in the region near the caustics but also in the region close 
to the planet-primary axis between the caustics. However, due to the 
difference in the locations of the caustics between the two types of 
systems with the planet-primary separations (normalized by the angular 
Einstein ring radius) $u_{\rm p}>1.0$ and $u_{\rm p}<1.0$, the locations 
of the major deviation regions of the two systems are different each 
other.  For systems with $u_{\rm p}>1.0$, the major deviation vectors
have orientations, in most cases, pointing towards the planet, while 
they point away from the planet for systems with $u_{\rm p}<1.0$.  
The major deviation region is surrounded by the region of moderate 
deviations, within which the orientation of $\Delta\deltavec$ is reversed 
compared to the orientation in the major deviation region.  We also find 
that the astrometric deviation is closely correlated with the photometric 
one as discussed by Safizadeh et al.  The astrometric deviation increases 
as the photometric deviation increases and $\Delta\deltavec$ is directed 
towards the planet when the light curve has positive deviation and vice 
versa.  We also present excess centroid shift maps for lens systems with 
various values of the planetary separation, planet/primary mass ratio, 
and source size to show the changes in the pattern of $\Delta\deltavec$ 
with these parameters.
\end{abstract}


\section{Introduction}
Although microlensing was originally proposed by Paczy\'nski (1986) as a 
method to detect Massive Astrophysical Compact Halo Objects (MACHOs), it 
can also be used in various other fields of astronomy. One important 
application is searching for extra-solar planets. Photometrically, planets 
can be found via the distortions they create in the lensing light curve 
compared to the smooth and symmetric one of a single lens event (Mao \& 
Paczy\'nski 1991; Gould \& Loeb 1992; Bolatto \& Falco 1994; Wambsganss 
1997). Due to the short duration (${\cal O}$ hour -- ${\cal O}$ day) of 
the planet-induced deviations, it is difficult to detect them from survey 
type experiments (e.g., OGLE: Udalski et al.\ 1993; MACHO: Alcock et al.\ 
1993; EROS: Aubourg et al.\ 1993; DUO: Alard \& Guibert 1997). However, 
with sufficiently frequent and accurate observations from follow-up 
monitoring of ongoing events alerted by the survey experiments, it is 
possible to detect such perturbations and characterize the mass ratio and 
the projected separation of the planet. Currently, several groups are 
carrying out such observations (MPS: Rhie et al.\ 2000; PLANET: Albrow et 
al.\ 1998; MOA: Bond 2000).

It was recently shown by Safizadeh, Dalal \& Griest (1999) that planets 
can also be detected from astrometric follow-up observations of lensing 
events by using the next generation of high-precision  interferometers 
such as the {\it Space Interferometry Mission} (SIM, Unwin et al.\ 1997) 
and those to be mounted on the Keck (Colavita et al.\ 1998) and the VLT 
(Mariotti et al.\ 1998).  Precise astrometry using these interferometers 
will permit measurements of the displacements in the source star image 
center of light with respect to its un-lensed position, $\deltavec$ 
(Miyamoto \& Yoshii 1995; H\o\hskip-1pt g, Novikov, \& Polnarev 1995; 
Walker 1995; Boden, Shao, \& Van Buren 1998; Miralda-Escud\'e 1996; 
Paczy\'nski 1998). The trajectory of the light centroid shifts (astrometric 
curve) caused by a single point mass is an ellipse (Walker 1995; Jeong, 
Han \& Park 1999). A planet can be detected because it can distort the 
elliptical astrometric curve of the single lens event, which is analogous 
to the deviation in the light curve.

In this paper, we investigate the properties of the planet-induced deviations 
in the astrometric curves.  For this, we construct a vector field map of 
the excess centroid shift $\Delta\deltavec$, which represents the difference 
of $\deltavec$ from that expected in the absence of the planet as a function 
of source positions.  With the map, we examine in detail the correlations 
between the astrometric and photometric deviations, which were briefly 
discussed by Safizadeh et al.\ (1992).

\section{Basics of Microlensing}
If a source located at $\thetavec_{\rm S}$ is gravitationally lensed by a 
coplanar $N$ point-mass lens system, where the individual masses and locations 
are $m_j$ and $\thetavec_{{\rm L},j}$, the positions of the resulting images 
$\thetavec$ are obtained by solving the lens equation of the form
\begin{equation}
\thetavec_{\rm S}=\thetavec - {\theta_{\rm E}^2\over m}
          \sum_{j=1}^{N}m_{j} { \thetavec-\thetavec_{{\rm L},j}\over 
	  {\left\vert \thetavec-\thetavec_{{\rm L},j}\right\vert}^2},
\end{equation}
where $m=\sum_{j=1}^N m_j$ is the total mass of the lens system and 
$\theta_{\rm E}$ is the angular Einstein ring radius.  The Einstein 
ring radius is related to the physical parameters of the lens system by
\begin{equation}
\theta_{\rm E} = \sqrt{4Gm\over c^2} 
\left({1\over D_{\rm OL}}-{1\over D_{\rm OS}}\right)^{1/2},
\end{equation}
where $D_{\rm OL}$ and $D_{\rm OS}$ represent the distances to the lens and 
source from the observer, respectively.

For the case of a single point-mass lens ($N=1$), there are two solutions 
(and thus the same number of images) of the lens equation and the resulting 
total magnification and the centroid shift vector take the simple forms
\begin{equation}
A = {u^2+2\over u\sqrt{u^2+4}},
\end{equation}
\begin{equation}
\deltavec = {\uvec\over u^2+2} \theta_{\rm E},
\end{equation}
where $\uvec=(\thetavec_{\rm S}-\thetavec_{\rm L})/\theta_{\rm E}$ is the 
dimensionless lens-source separation vector normalized by $\theta_{\rm E}$.  
The separation vector is related to the single lensing parameters by
\begin{equation}
\uvec = \left( {t-t_0\over t_{\rm E}}\right)\ \hat{\xvec}
        \ \pm\ \beta\ \hat{\yvec},
\end{equation}
where $t_{\rm E}$ is the Einstein ring radius crossing time (Einstein time 
scale), $\beta$ is the closest lens-source separation (impact parameter), 
$t_0$ is the time at the moment of the closest approach, and the unit 
vectors $\hat{\xvec}$ and $\hat{\yvec}$ are parallel and normal to the 
direction of the lens-source transverse motion, respectively.  Note that 
the sign `$\pm$' is used because $\beta$ is positive definitive.

The lens system with a planet is described by the formalism of the binary 
lens system (i.e., $N=2$) with a very low mass companion.  For this case, 
the lens equation becomes a fifth degree polynomial equation (Witt \& Mao 
1995) and there are three or five solutions (images) depending on the source 
position with respect to the lenses.  The magnifications of the individual 
images are given by the Jacobian of the transformation (1) evaluated at the 
image position, i.e.\
\begin{equation}
A_i = \left({1\over \vert {\rm det}\ J\vert} \right)_{\thetavec=\thetavec_i};
\qquad 
	{\rm det}\ J = \left\vert{\partial\thetavec_{\rm S}
        \over \partial\thetavec}
\right\vert.
\end{equation}
Then, the total magnification is given by the sum of the magnifications 
of the individual images, i.e.\ $A=\sum_{i=1}^{N_I} A_i$, where $N_I$ is 
the total number of images.  Since the position of the image centroid,
$\thetavec_{\rm c}$, equals the magnification weighted mean position of 
the individual images, the source image centroid shift with respect to the 
unlensed source position is computed by
\begin{equation}
\deltavec=\thetavec_{\rm c}-\thetavec_{\rm S};\ \ 
\thetavec_{\rm c}={\sum_{i}^{N_I} A_i \thetavec_i \over A}.
\end{equation}

The fundamental difference in the geometry of a binary lens system from 
that of a single point-mass lens is the formation of caustics.  The 
caustics are the set of source positions at which the magnification of 
a point source diverges in the geometric optics limit, i.e.\ 
${\rm det}\ J=0$.  Hence, the most significant planet-induced deviations 
both in the light and astrometric curves occur when the source passes the 
region close to the caustics. The set of caustics form closed curves.  
The maximum size of the caustics occurs when the planet is located in the 
lensing zone of $0.6 \lesssim u_{\rm p}\lesssim 1.6$ (Gould \& Loeb 1992), 
where $u_{\rm p}$ is the projected primary-planet separation in units of 
$\theta_{\rm E}$ (planetary separation).

\section{Maps of the Excess Centroid Shifts}

We investigate the properties of planet-induced deviations in the astrometric 
curves by constructing maps of excess centroid shifts.  The excess centroid 
shift is defined by 
\begin{equation}
\Delta\deltavec = \deltavec_{\rm p} - \deltavec,
\end{equation}
where $\deltavec_{\rm p}$ and $\deltavec$ represent the centroid shifts with 
the presence and absence of the planet, respectively.  Therefore, the excess 
centroid shift vector tells us how much the astrometric curve of an event 
caused by a lens system with a planet deviates from the elliptical one of a 
single point-mass event and in which direction the deviation is directed. 
The excess centroid shift map is analogous to the excess magnification map 
that was often used to estimate the probability of photometric planet 
detections (Gould \& Loeb 1992; Gaudi \& Gould 1997; Gaudi \& Sackett 2000).  
The excess magnification is defined by 
\begin{equation}
\epsilon = {A_{\rm p}-A\over A},
\end{equation}
where $A_{\rm p}$ and $A$ are the magnifications with and without the planet,
respectively.  We note, however, that $\Delta\deltavec$ is a vector quantity, 
while $\epsilon$ is a scalar quantity. Hence the excess centroid shift map 
is a vector field map.  With the map, one can obtain an overview of the 
astrometric behaviors of planet-induced perturbations of all events caused 
by a lens system, without testing their individual astrometric curves 
(Han 2001).  For the construction of the map, the excess centroid shift on 
each source position is determined by calculating $\deltavec$ and 
$\deltavec_{\rm p}$.  We calculate $\deltavec_{\rm p}$ by numerically 
solving the binary lens equation, i.e.\ by solving the fifth degree 
polynomial equation (Witt \& Mao 1995).

\section{The Properties of Excess Centroid Shifts}
As will be shown in the following subsections, the properties of astrometric 
perturbations are greatly different between the systems with $u_{\rm p}>1.0$ 
and $u_{\rm p}<1.0$.  Hence, we separately investigate the properties of 
$\Delta\deltavec$ for these two types of systems.

\subsection{Planets with $u_{\rm p}>1.0$}
For this type of lens system, there exist two disconnected sets of caustics. 
One is located near the center of mass (central caustic) and the other 
caustic (planetary caustic) is located between the primary and the planet
(Griest \& Safizadeh 1998; Bozza 1999).

In the upper panel of Figure 1, we present the excess centroid shift map 
of an example lens system composed of a planet with $u_{\rm p}>1.0$.  The 
grey-scales in the map are used to represent the regions of significant 
deviations with $\Delta\delta \geq 0.1\theta_{\rm E}$ (dark shaded region) 
and moderate deviations with $\Delta\delta \geq 0.03\theta_{\rm E}$ (light 
shaded region).  The angular Einstein ring radius of a Galactic bulge event 
caused by a stellar mass lens with $m\sim 0.3\ M_\odot$ located midway
between the observer and the source, i.e.\ $D_{\rm OL}/D_{\rm OS}=0.5$, 
is $\theta_{\rm E}\sim 550$ $\mu$-arcsec.  For this event, the deviations 
in these regions respectively are $\Delta\delta \gtrsim 50$ $\mu$-arcsec 
and $\Delta\delta \gtrsim 15$ $\mu$-arcsec.  In Figure 2 and 3, we present 
the light and astrometric curves of the events resulting from the source 
trajectories marked in the map (long straight lines with arrows). The 
assumed planetary separation is $u_{\rm p}=1.2$ and the planet/primary 
mass ratio is $q=10^{-3}$, which is roughly equivalent to the mass ratio 
between the Jupiter and the Sun.  To show how the astrometric deviation 
is correlated to the photometric one, we also present the excess 
magnification map (the contour map in the second panel of Fig.\ 1) and the 
curves of $\epsilon$ and $\Delta\deltavec$ around the time and near the 
region of significant deviations (Figure 4). The contours of the excess 
magnification map are drawn at the levels of $\epsilon=-10\%$, $-5\%$, 
5\%, and 10\% and the regions of positive excesses are distinguished by 
grey-scales.

From the analysis of the figures, we find the following properties of 
$\Delta\deltavec$ induced by planets with separations from the
primary larger than $\theta_{\rm E}$.
\begin{enumerate}
\item
The most significant astrometric deviations ($\Delta\delta\gtrsim 0.1
\theta_{\rm E}$) occur in the region near the caustics and also in the 
region close to the primary-planet axis between the two caustics.  Since the 
planetary caustic for this system is located on the planet side with 
respect to the center of mass, the major deviation region is also located 
on the planet side. The biggest deviations occur when the source crosses 
the caustics (e.g., the event caused by the source trajectory 3). For 
this case, the resulting astrometric curve exhibits a sharp spike due 
to the sudden shift of the centroid (see the corresponding $\deltavec$ 
and $\Delta\deltavec$ of the event in Fig.\ 3 and 4).\footnote{For more 
detailed discussion about the behavior of $\Delta\deltavec$ for caustic 
crossing events, see \S\ 3.3 of Han (2000).}  Even for non-caustic crossing 
events, however, significant deviations can also occur if the source 
trajectory passes through the primary-planet axis region between the central 
and planetary caustics. For this case, the deviation is smooth as shown in 
Fig.\ 3 and the right panels of Fig.\ 4.

\item
The deviation vectors in most of the major deviation region are directed 
towards the planet along the primary-planet axis. As a result, a majority 
of the events caused by a planetary lens system with $u_{\rm p}>1.0$ will 
have major deviations oriented towards the planet. The only exception occurs 
if the source passes through the small region on the opposite side of the 
primary lens (e.g., the event caused by the source trajectory 6), within 
which the major deviation vector is directed away from the planet.

\item
As the distance from the primary-planet axis increases, the amount of 
deviation decreases.  In addition, the direction of $\Delta\deltavec$ is 
reversed compared to that of $\Delta\deltavec$ in the major deviation region. 
Since the major deviation region is surrounded by the region of moderate 
deviations, the astrometric curve of a typical event with significant 
deviations will have smaller deviations with an opposite orientation to 
that of the major deviation vectors before and after the major deviation.
We note that planet-induced photometric deviations also exhibit a similar 
property, i.e.\ having smaller deviations with an opposite sign of $\epsilon$ 
around the major deviation (Gaudi \& Gould 1992).

\item
The photometric and astrometric deviations are closely correlated, as 
discussed by Safizadeh et al.\ (1999).  From the comparison of the sizes 
between the photometric and the corresponding astrometric deviations, one 
finds that the amount of $\Delta\delta$ increases as $\left\vert \epsilon
\right\vert$ increases. From the comparison of the sign of $\epsilon$ and 
the direction of $\Delta\deltavec$, one also finds that $\Delta\deltavec$ 
is directed towards the planet when $\epsilon>0$ (positive deviation) and 
vice versa.  Since the major astrometric deviation vectors of the events 
caused by this type of lens system are usually directed towards the planet, 
the light curves of these events are more likely to have positive major 
deviations (see Fig.\ 2 and the right panels of Fig.\ 4).  The location of 
the major astrometric deviation regions and the orientations of $\Delta
\deltavec$ in this region can be understood from the correlations between 
$\Delta\deltavec$ and $\epsilon$.  The most significant photometric 
deviations occur when the source is located near the primary-planet axis
because at this moment the images are located near the axis and approaches 
most closely the planet (Gould \& Loeb 1992).  As a result, the region of 
major astrometric deviations is located along the axis and $\Delta\deltavec$ 
in this region is directed also along the axis.  In addition, the similarity 
in the properties of moderate size deviations between the astrometric and 
photometric perturbations around the major deviations mentioned above can 
also be understood from the correlations between $\Delta\deltavec$ and 
$\epsilon$.
\end{enumerate}

\subsection{Planets with $u_{\rm p}<1.0$}
The lens system composed of a planet with $u_{\rm p}<1.0$ also has a single 
central caustic, but it has {\it two} planetary caustics, which are located 
on the {\it opposite} side of the planet with respect to the center of mass.
Unlike the planetary caustic of the planetary lens system with $u_{\rm p}>1.0$,
the planetary caustics of this lens system are {\it not} located on the 
primary-planet axis, although they are symmetric with respect to the axis 
(Griest \& Safizadeh 1998).

In the upper panel of Figure 5, we present the map of $\Delta\deltavec$ 
for a planetary lens system with $u_{\rm p}=0.8$ and $q=10^{-3}$ and it is 
compared to the excess magnification map presented in the second panel. 
Also presented in Figure 6 and 7 are the light and astrometric curves of 
several example events resulting from the source trajectories marked in 
the upper panel of Fig.\ 5. In Figure 8, we present the photometric and 
astrometric deviations around the time and near the region of significant 
deviations for these events. From the investigation of the properties of 
$\Delta\deltavec$ induced by the planet with $u_{\rm p}<1.0$ and comparison 
to those of the deviation vectors induced by the planet with $u_{\rm p}>1.0$, 
we find the following similarities and differences.
\begin{enumerate}
\item
For this system, the most significant deviations also occur near the 
caustic regions and the region close to the primary-planet axis between 
the two types of caustics. However, since the planetary caustics are located 
on the opposite side of the planet with respect to the center of mass, the 
significant deviation regions are  also located on the opposite side of 
the planet.  

\item
In addition, besides the small areas around the planetary caustics, 
$\Delta \deltavec$ in most of the major deviation region are directed 
away from the planet along the primary-planet axis 
(see Fig.\ 7 and right panels of Fig.\ 8).  

\item
Similar to the behaviors of $\Delta\deltavec$ induced by the planet with 
$u_{\rm p}>1.0$, the major deviation region is surrounded by the region 
of moderate deviations, within which the direction of $\Delta\deltavec$ 
is reversed to that of $\Delta\deltavec$ in the major deviation region.

\item
The relation between the astrometric and photometric deviations is also
similar to the case of the planetary system with $u_{\rm p}>1.0$; the amount 
of astrometric deviation increases as $\left\vert\epsilon \right\vert$ 
increases and its direction points towards the planet when the light 
curve has a positive deviation and vice versa. Because the major astrometric 
deviations induced by this type of planets are usually directed towards the 
opposite side of the planet, the light curves have negative deviations 
($\epsilon<0$) with the same probability.

\item
Another interesting finding is that most of the major planet-induced 
astrometric deviations point {\it away} from the opening of the astrometric 
ellipse regardless of the planetary separation (see Fig.\ 3 and 7).
\end{enumerate}

\section{Variations}
In this section, we investigate the variation of the excess centroid shift 
patterns with the planetary separation and the planet/primary mass ratio.
For this, we present excess centroid shift maps of lens systems with various 
values of $u_{\rm p}$ and $q$.  In addition, to examine the effect of finite 
source size on the pattern of $\Delta\deltavec$, we present maps of an 
example lens system expected for different source sizes.

In Figure 9, we present the locations of significant astrometric deviation
regions for lens systems with various values of $u_{\rm p}$. In the following
Figures 10 and 11, we present the vector field maps of $\Delta\deltavec$
of the corresponding lens systems in the regions around the significant
deviations.  The tested lens systems have a common mass ratio of $q=10^{-3}$ 
and a variety of planetary separations of $u_{\rm p}=2.0$, 1.7, 1.5, 1.3 
(whose maps are presented in Fig.\ 10), 1.0, 0.9, 0.7, and 0.5 (whose maps 
are presented in Fig.\ 11).  One finds that similar to photometric deviations 
(Gould \& Loeb), the size of significant deviation regions is maximized 
when $u_{\rm p}\sim 1.0$ and becomes smaller as the separation increases or 
decreases from this value.  We note, however, that the general patterns of 
$\Delta\deltavec$ described in \S\ 4.1 and \S\ 4.2 still apply.

In Figure 12 and 13, we present the excess centroid shift maps of lens 
systems with various values of planet/primary mass ratios.  The tested 
systems have a common planetary separation of $u_{\rm p}=1.3$ and various 
mass ratios of $q=10^{-2}$, $5\times 10^{-3}$, $3\times 10^{-3}$, $10^{-3}$ 
(whose maps are presented in Fig.\ 12), $5\times 10^{-4}$, $10^{-3}$, 
$5\times 10^{-5}$, and $10^{-5}$ (whose maps are presented in Fig.\ 13). One 
finds that the location of the major astrometric deviation region does not 
vary with mass ratio, but the size of the region decreases rapidly as $q$ 
decreases.  This property is also similar to that of photometric deviations.

In Figure 14, we presented the excess centroid shift maps of a common 
lens system (with $u_{\rm p}=1.2$ and $q=10^{-3}$) but with different 
source sizes.  Considering the finite source effect, the position of the 
image centroid is given by 
\begin{equation}
\thetavec_{\rm c}=
{
\int\ I(\rvec)\ A_{\rm p}(\rvec)\ \theta_{\rm c,p}(\rvec)\ d^2 r
\over 
\int\ I(\rvec)\ A_{\rm p}(\rvec)\ d^2 r
}
\end{equation}
where $I(\rvec)$ is the surface intensity distribution of the source star
and $\rvec$ represents the displacement vector of an area element on the 
source star surface with respect to the lens.  For the construction of 
the maps, we assume the distribution of the source star surface intensity 
is uniform.  We note that $\Delta\deltavec$ in the maps are the deviations 
from the centroid shifts of the single primary lens which are also affected 
by the finite source effect.  When a single lens event is affected by the 
finite source effect, the resulting centroid shift trajectory is no longer 
an ellipse (Mao \& Witt 1998).  We test three different source sizes of 
$\varrho_\ast=0.005\theta_{\rm E}$, $0.01\theta_{\rm E}$, and 
$0.05\theta_{\rm E}$.  From the figure, one finds that the detailed 
structure of $\Delta\deltavec$ is smeared out with the increasing source 
size, as pointed out by Safizadeh et al.\ (1998).  For example planetary 
microlensing astrometric curves affected by the finite source effect, 
see Fig.\ 3 of Safizadeh et al.\ (1998).

\section{Conclusion}
We investigate the astrometric properties of planet-induced deviations 
in the trajectory of the microlensed source star's centroid shift motion.
For this purpose, we construct maps of excess centroid shifts from which 
one can obtain an overview about the behaviors of $\Delta\deltavec$ of 
all events without testing their individual centroid shift trajectories.  
From the analysis of the maps for various types of lens systems, we find 
that major astrometric deviations occur not only in the region around caustic 
but also  in the region close to the primary-planet axis between caustics.  
The region of major deviations is surrounded by the region of moderate 
deviations in which the orientation of $\Delta\deltavec$ is reversed 
compared to that in the major deviation region.  Due to the difference in 
the locations of the caustics between the planetary lens systems with 
$u_{\rm p}>1.0$ and $u_{\rm p}<1.0$, the locations of the major deviations 
and the orientations of $\Delta\deltavec$ within these regions of the two 
types of lens systems are different from each other.  From the investigation 
of the correlations between the astrometric and photometric deviations, we 
find that both the size and the direction of the astrometric deviation are 
closely related to the magnitude and sign of the photometric deviation, as 
discussed by Safizadeh et al.\ (1998).  By investigating the variation of 
the pattern of $\Delta\deltavec$ with  the planetary separation, 
planet/primary mass ratio, and source size, we find the astrometric and 
photometric deviations both have similar dependence upon these parameters.

We would like to thank K.\ Griest for making useful comments about the paper.
This work was supported by a grant (Basic Research Fund) of the Korea 
Science \& Engineering Foundation (KOSEF).

{}

\clearpage

\begin{figure*}
\epsfysize=15cm
\centerline{\epsfbox{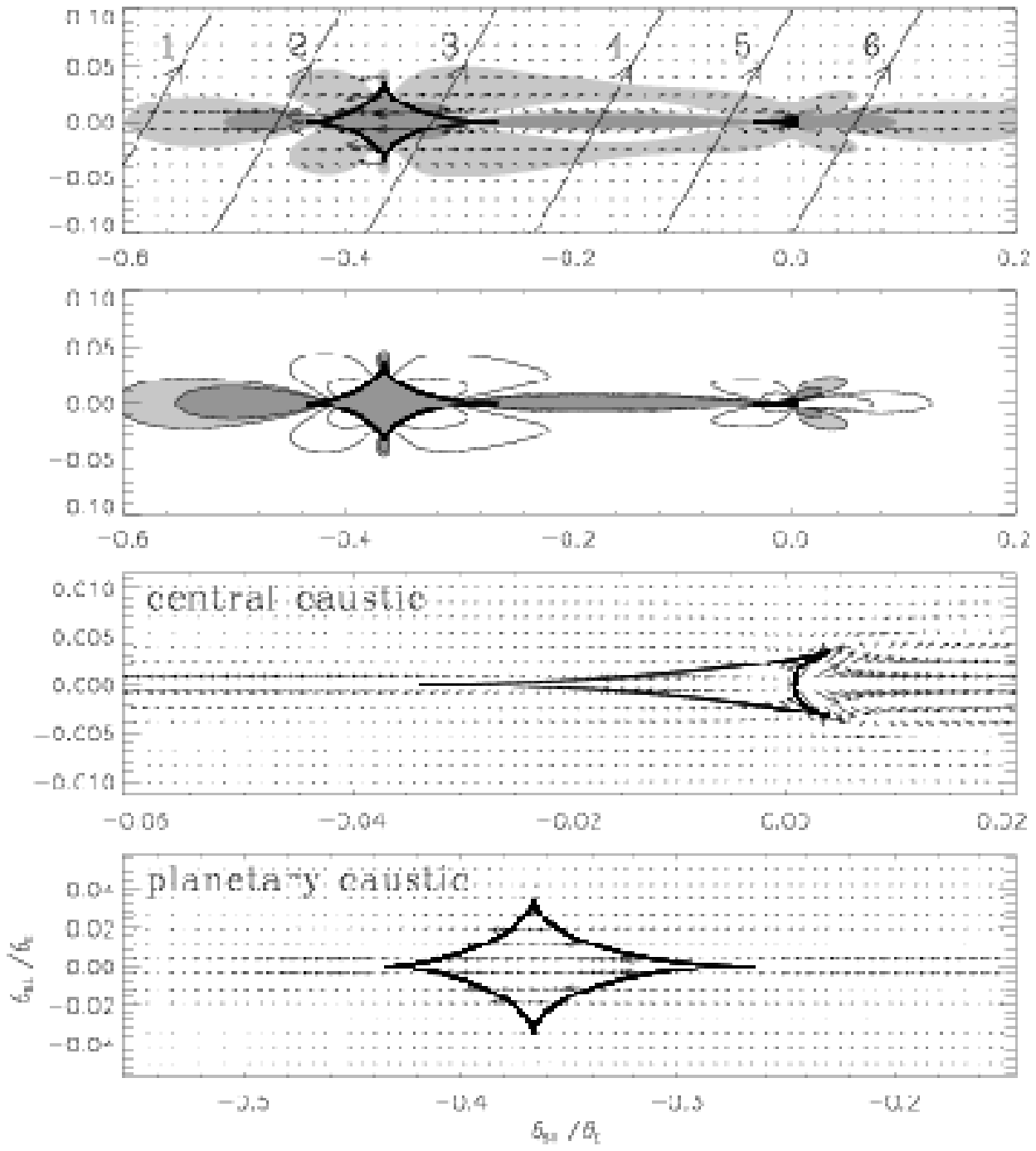}}
\caption{
The vector field map of the excess centroid shift $\Delta\deltavec$ of an
example planetary lens system with a planetary separation $u_{\rm p}=1.2$
nd a mass ratio $q=10^{-3}$.  The positions are selected so that the center
of mass is at the origin and $(\theta_{{\rm S}\parallel},\theta_{{\rm S}\bot})$
represent the components of the source position that are parallel and normal
to the primary-planet axis. Both the primary and the planet are located on
the $\theta_{{\rm S}\parallel}$ axis and the planet is to the left. The two
closed figures drawn by thick solid lines represent the caustics. Grey-scales
are used to represent the regions of significant deviations with $\Delta\delta
\geq 0.1\theta_{\rm E}$ (dark shaded region) and moderate deviations with
$\Delta\delta\geq 0.03 \theta_{\rm E}$ (light shaded region). The long
straight lines with arrows represent the source trajectories of the events
whose resulting light and astrometric curves are presented in Fig.\ 2 and 3,
respectively.  Second panel: The contour map of the excess magnification for
the same lens system. The contours are drawn at the levels of $\epsilon=-10\%$,
$-5\%$, 5\%, and 10\%.  The region of positive excesses are distinguished by
grey-scales.  Lower two panels: Blow-ups of the excess centroid shift map in
the regions around the central (third panel) and planetary caustics (last
panel).
}
\end{figure*}

\begin{figure*}
\epsfysize=15cm
\centerline{\epsfbox{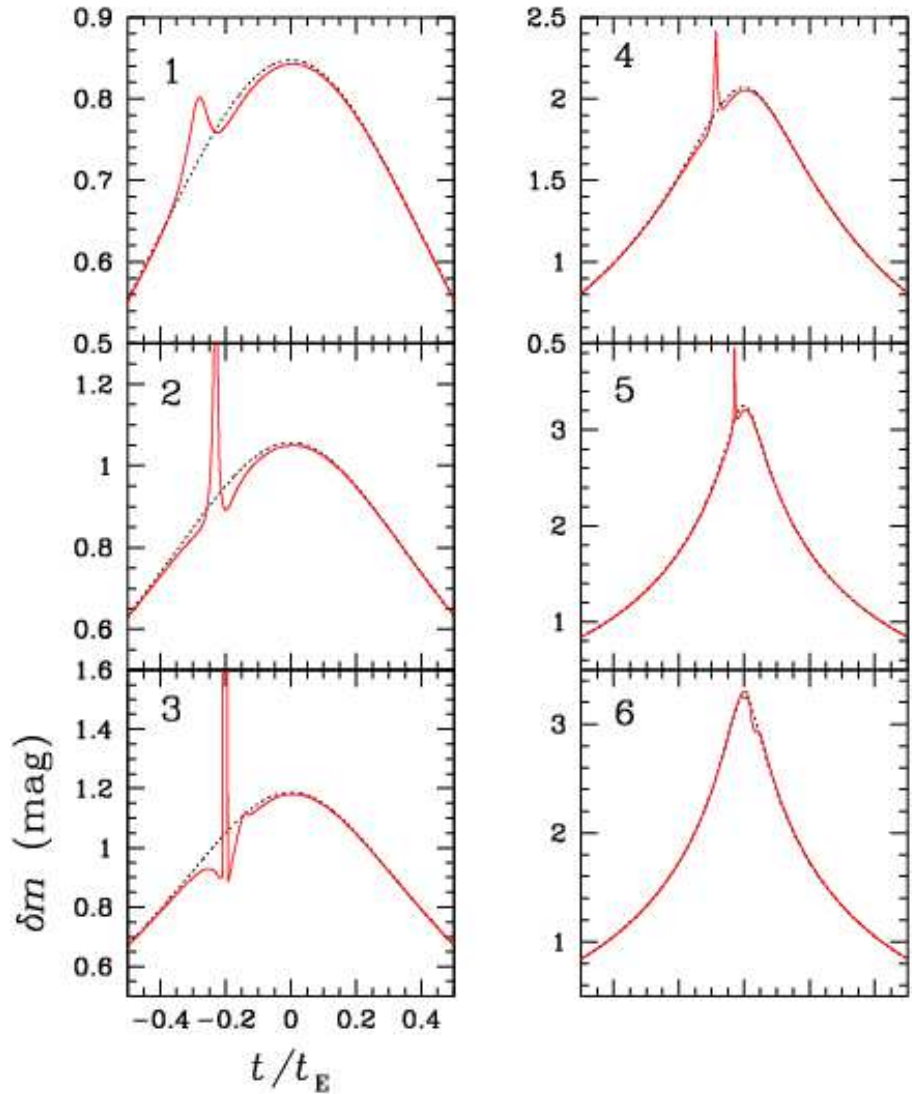}}
\caption{
The light curves of the events resulting from the source trajectories marked
in the upper panel of Fig.\ 1.  The number in each panel corresponds to the 
trajectory number marked in Fig.\ 1.  The dotted curves represent the single 
lens event light curves that are expected in the absence of the planet.  }
\end{figure*}

\begin{figure*}
\epsfysize=15cm
\centerline{\epsfbox{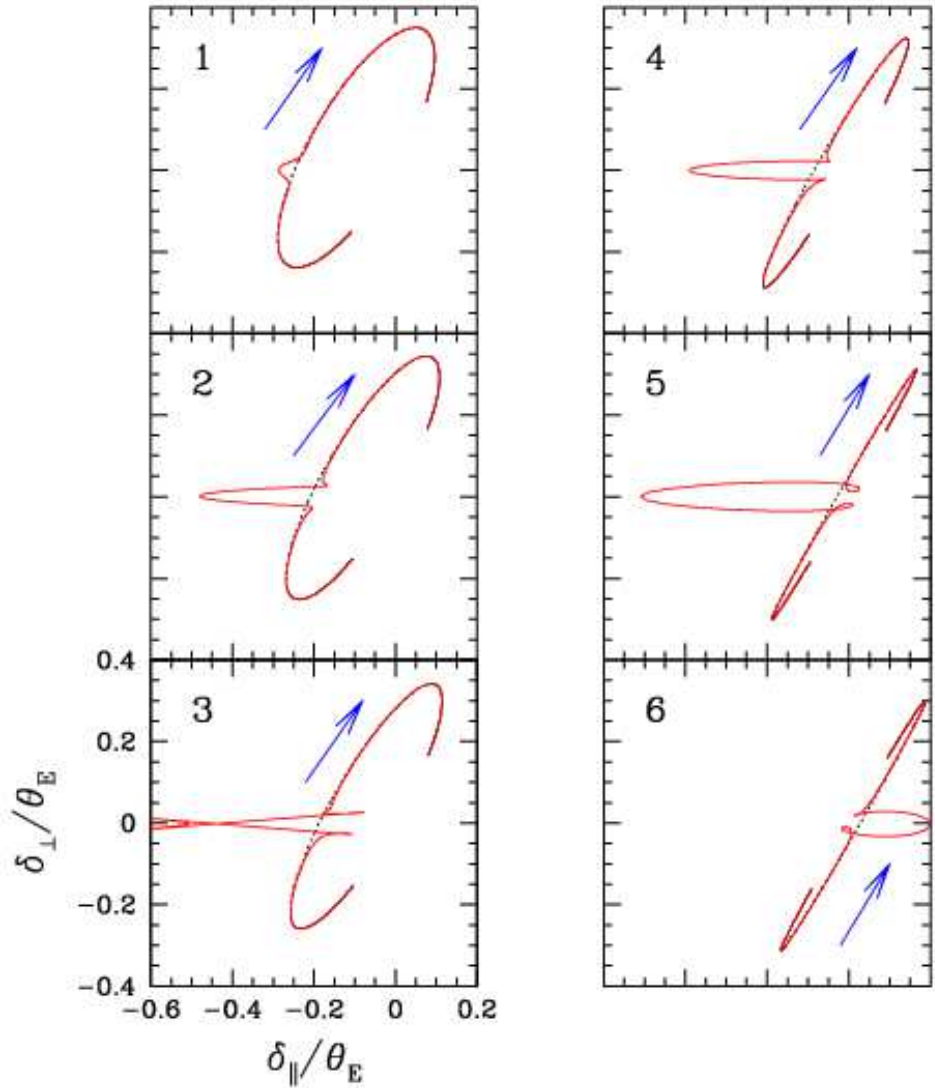}}
\caption{
The astrometric curves of the events resulting from the source trajectories
marked in the upper panel of Fig.\ 1. The dotted curves represent the 
astrometric curves expected in the absence of the planet. $(\delta_\parallel,
\delta_\bot)$ are the components of $\deltavec$ that are parallel and normal 
to the primary-planet axis and the arrow represents the direction of the 
centroid motion.  }
\end{figure*}

\begin{figure*}
\epsfysize=15cm
\centerline{\epsfbox{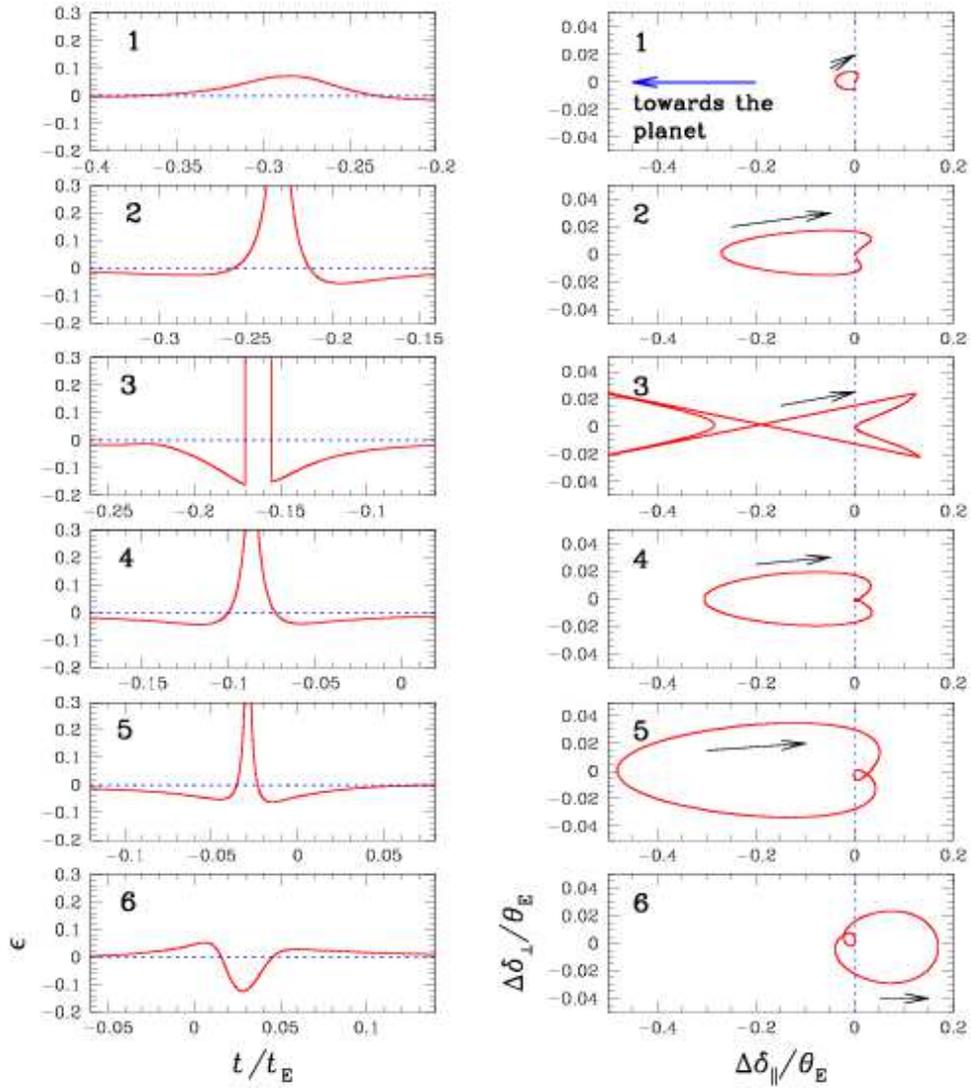}}
\caption{
The excess magnification $\epsilon$ (left panels) and the excess centroid 
shift vector $\Delta\deltavec$ (right panels) around the time and near the 
region of significant deviations for the events resulting from the source 
trajectories marked in Fig.\ 1. The number in each panel corresponds to the 
trajectory number. Note that the abscissa and ordinate of $\Delta\deltavec$ 
are arbitrarily scaled.  }
\end{figure*}

\begin{figure*}
\epsfysize=15cm
\centerline{\epsfbox{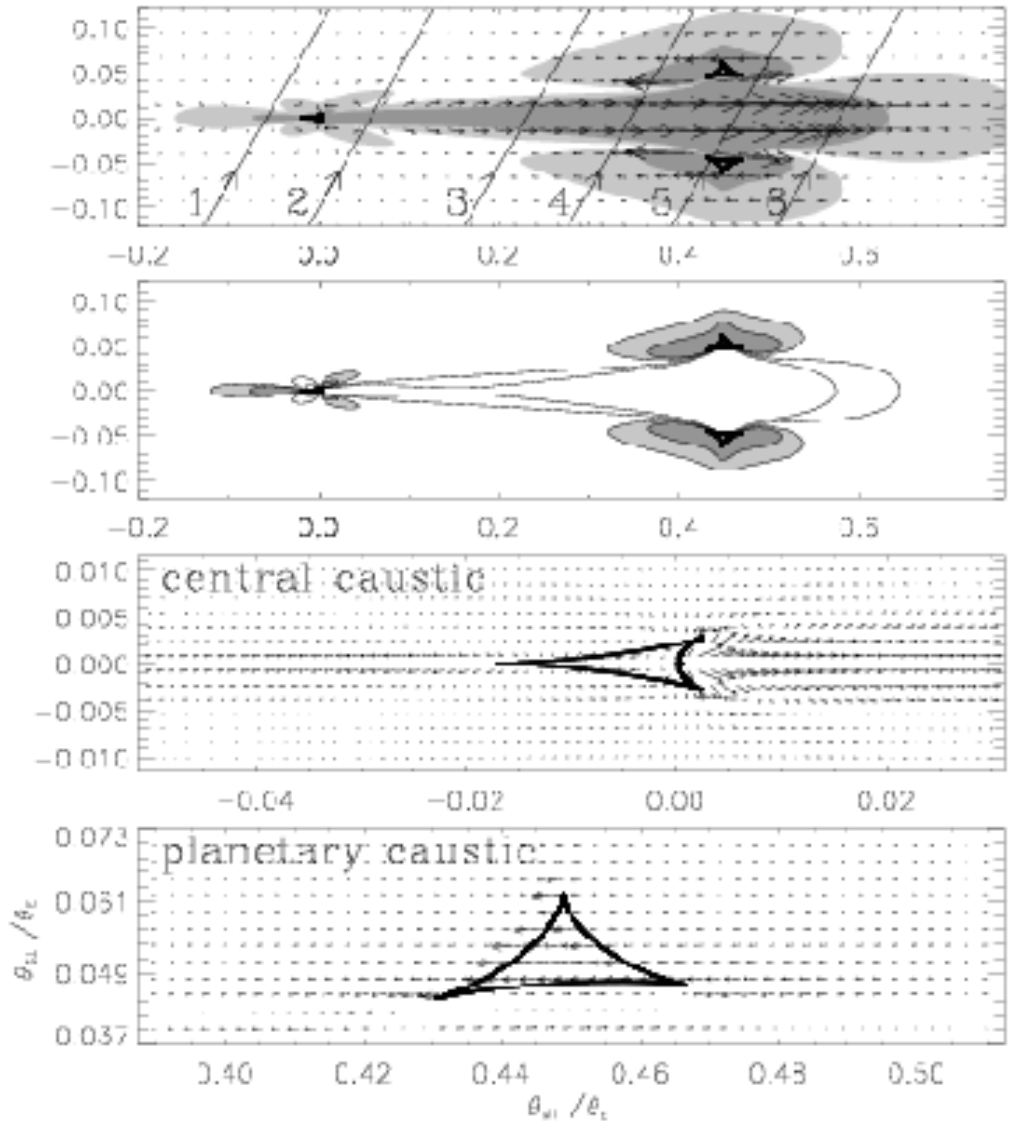}}
\caption{
The vector field map of $\Delta\deltavec$ (upper panel) and the contour map 
of $\epsilon$ (second panel) of an example planetary lens system with 
$u_{\rm p}=0.8$ and $q=10^{-3}$.  The lower two panels are the blow-ups of 
the the excess centroid shift map in the region around caustics.  Notations 
are same as in Fig.\ 1.  }
\end{figure*}

\begin{figure*}
\epsfysize=15cm
\centerline{\epsfbox{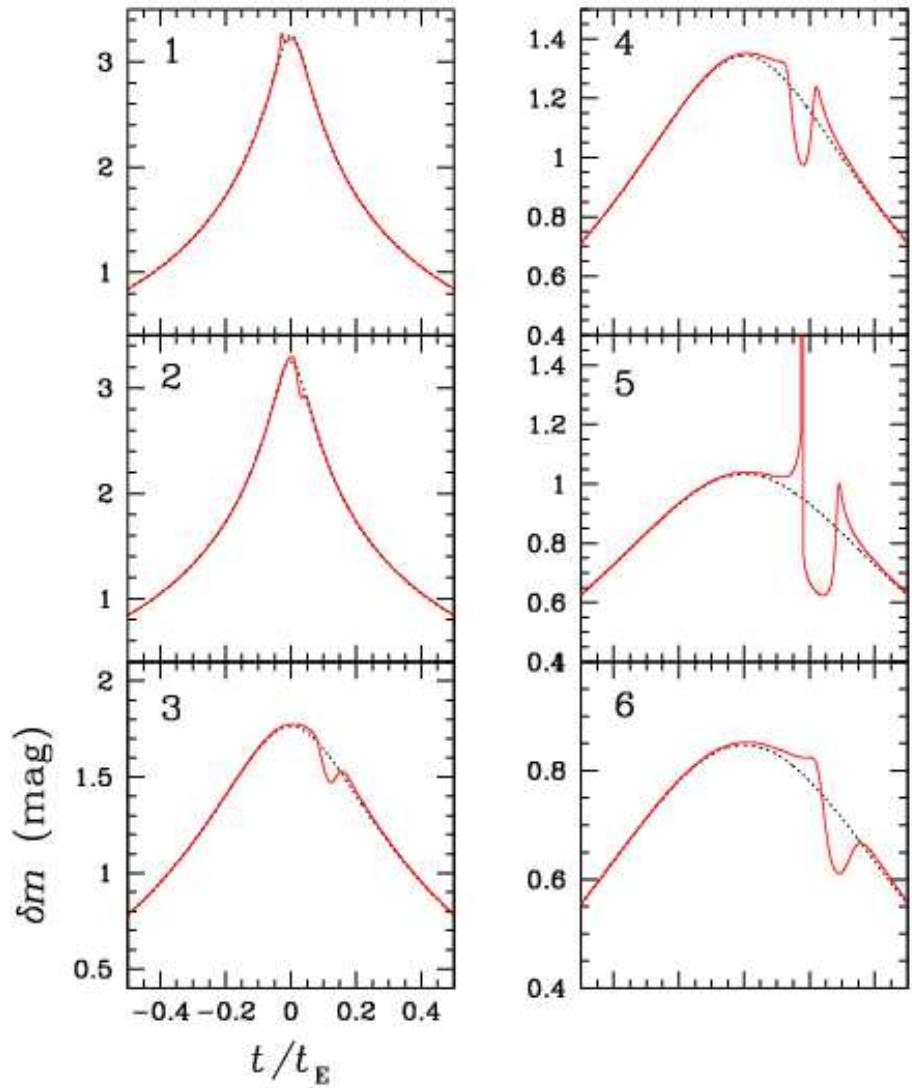}}
\caption{
The light curves of the events resulting from the source trajectories marked
in the upper panel of Fig.\ 5.  Notations are same as in Fig.\ 2.  }
\end{figure*}

\begin{figure*}
\epsfysize=15cm
\centerline{\epsfbox{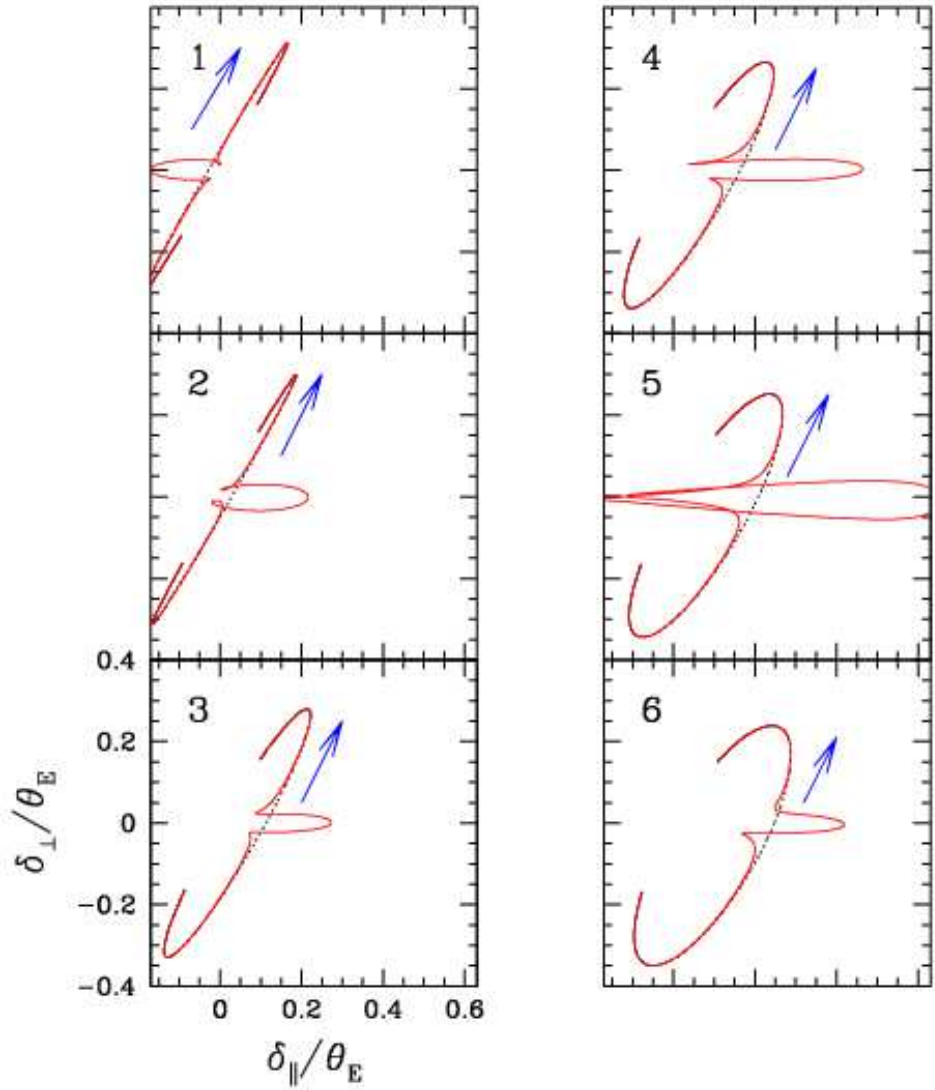}}
\caption{
The astrometric curves of the events resulting from the source trajectories
marked in the upper panel of Fig.\ 5. Notations are same as in Fig.\ 3.}
\end{figure*}

\begin{figure*}
\epsfysize=15cm
\centerline{\epsfbox{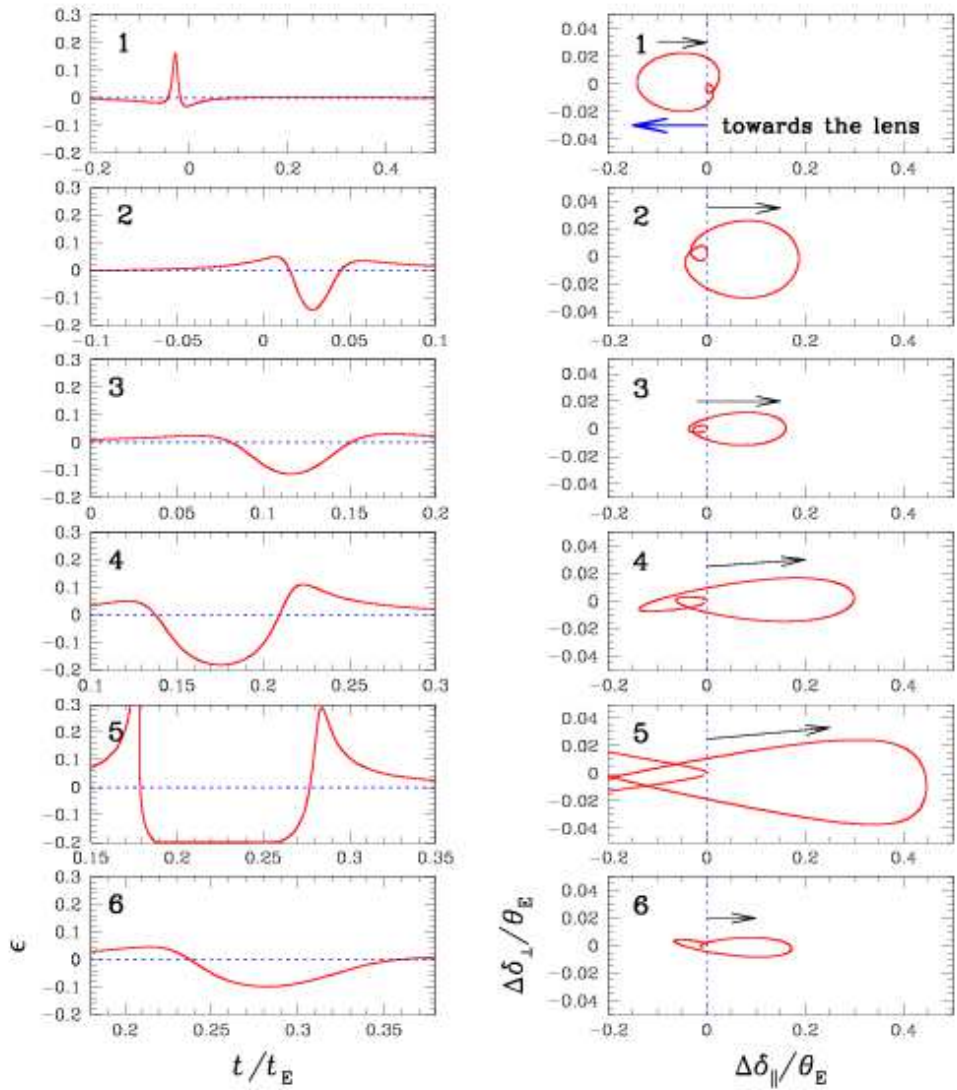}}
\caption{
The excess magnification $\epsilon$ (left panels) and the excess centroid
shift vector $\Delta\deltavec$ (right panels) for the events resulting 
from the source trajectories marked in Fig.\ 5. Notations are same as in 
Fig.\ 4.}
\end{figure*}

\begin{figure*}
\epsfysize=13cm
\centerline{\epsfbox{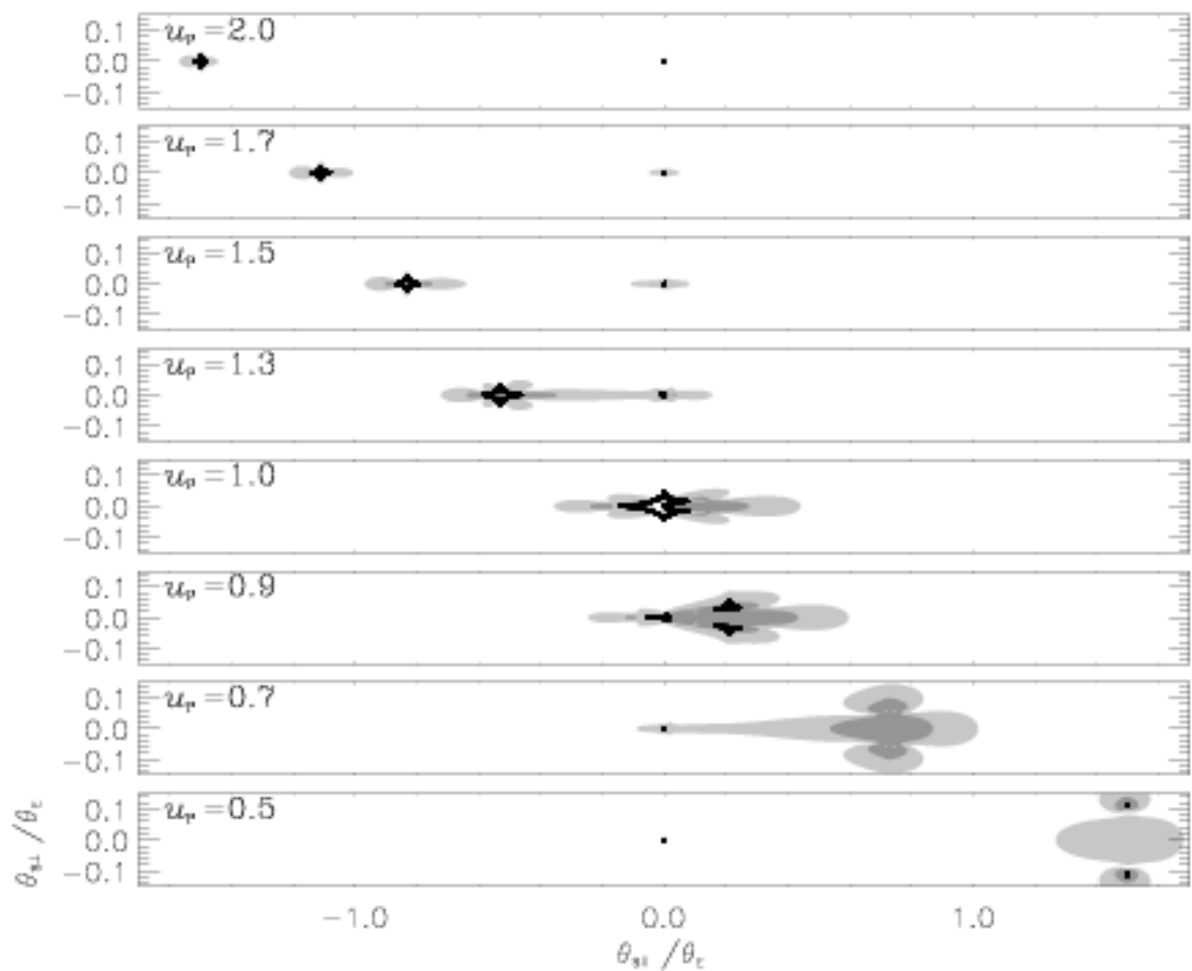}}
\caption{
Variation of the locations of significant astrometric deviation regions 
with planetary separation $u_{\rm p}$.  The light and dark shades are used 
to represent the regions with $\Delta\delta \geq 0.03\theta_{\rm E}$ and 
$\Delta\delta \geq 0.1\theta_{\rm E}$, respectively.  The lens systems have 
a common mass ratio of $q=10^{-3}$.  The vector field maps of $\Delta\deltavec$
of the individual lens systems are presented in the following Fig.\ 10 and 11.
}
\end{figure*}

\begin{figure*}
\epsfysize=17cm
\centerline{\epsfbox{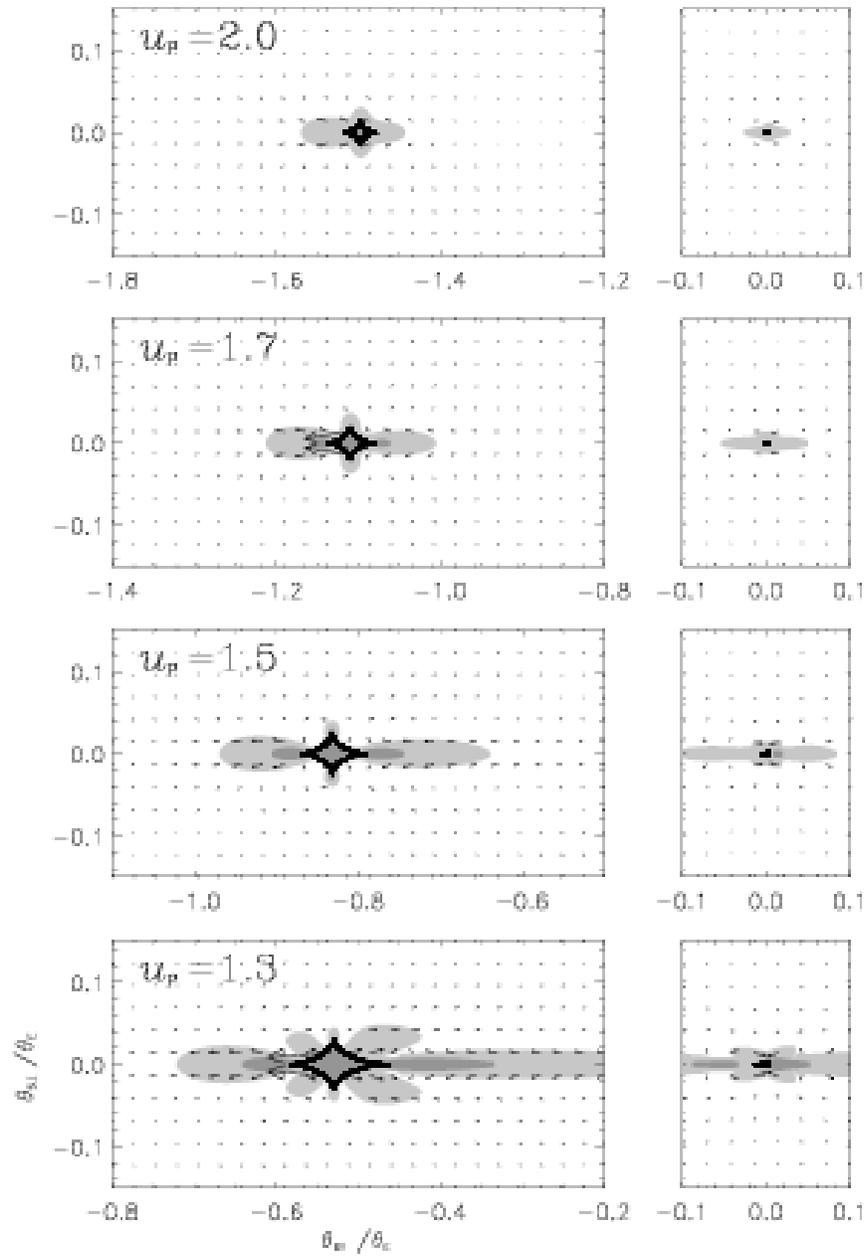}}
\caption{
Variation of the vector field map of $\Delta\deltavec$ with the planetary 
separation $u_{\rm p}$.  The lens systems have a common mass ratio of 
$q=10^{-3}$.  Notations are same as in the upper panel of Fig.\ 1.
}
\end{figure*}

\begin{figure*}
\epsfysize=17cm
\centerline{\epsfbox{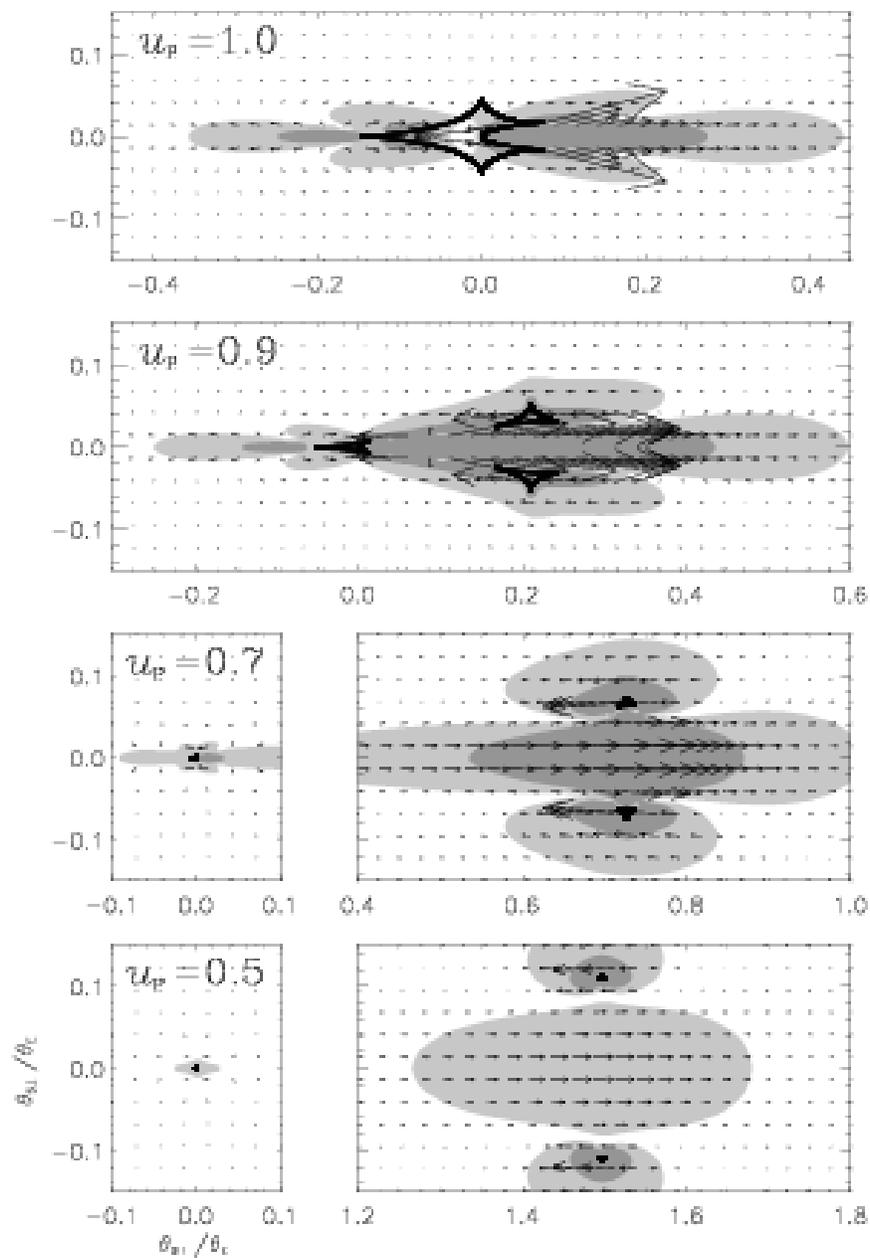}}
\caption{
Variation of the vector field map of $\Delta\deltavec$ with the planetary
separation $u_{\rm p}$.  The maps are similar to those presented in Fig.\ 10, 
but for lens systems with smaller planetary separations.
}
\end{figure*}

\begin{figure*}
\epsfysize=17cm
\centerline{\epsfbox{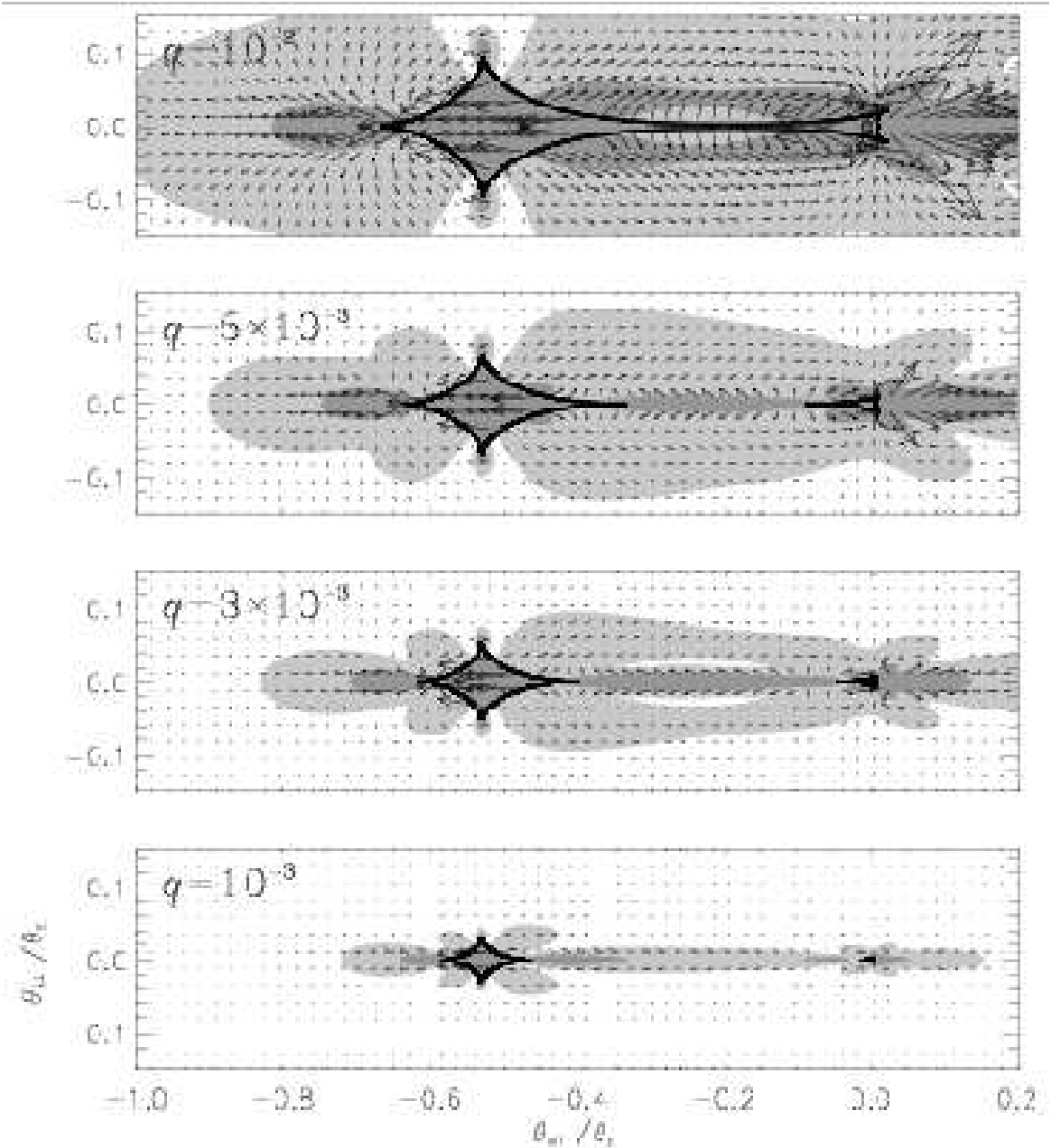}}
\caption{
Variation of the vector field map of $\Delta\deltavec$ with the 
planet/primary mass ratio $q$.  The lens systems have a common planetary 
separation of $u_{\rm p}=1.3$.  Notations are same as in the upper panel 
of Fig.\ 1.
}
\end{figure*}

\begin{figure*}
\epsfysize=15cm
\centerline{\epsfbox{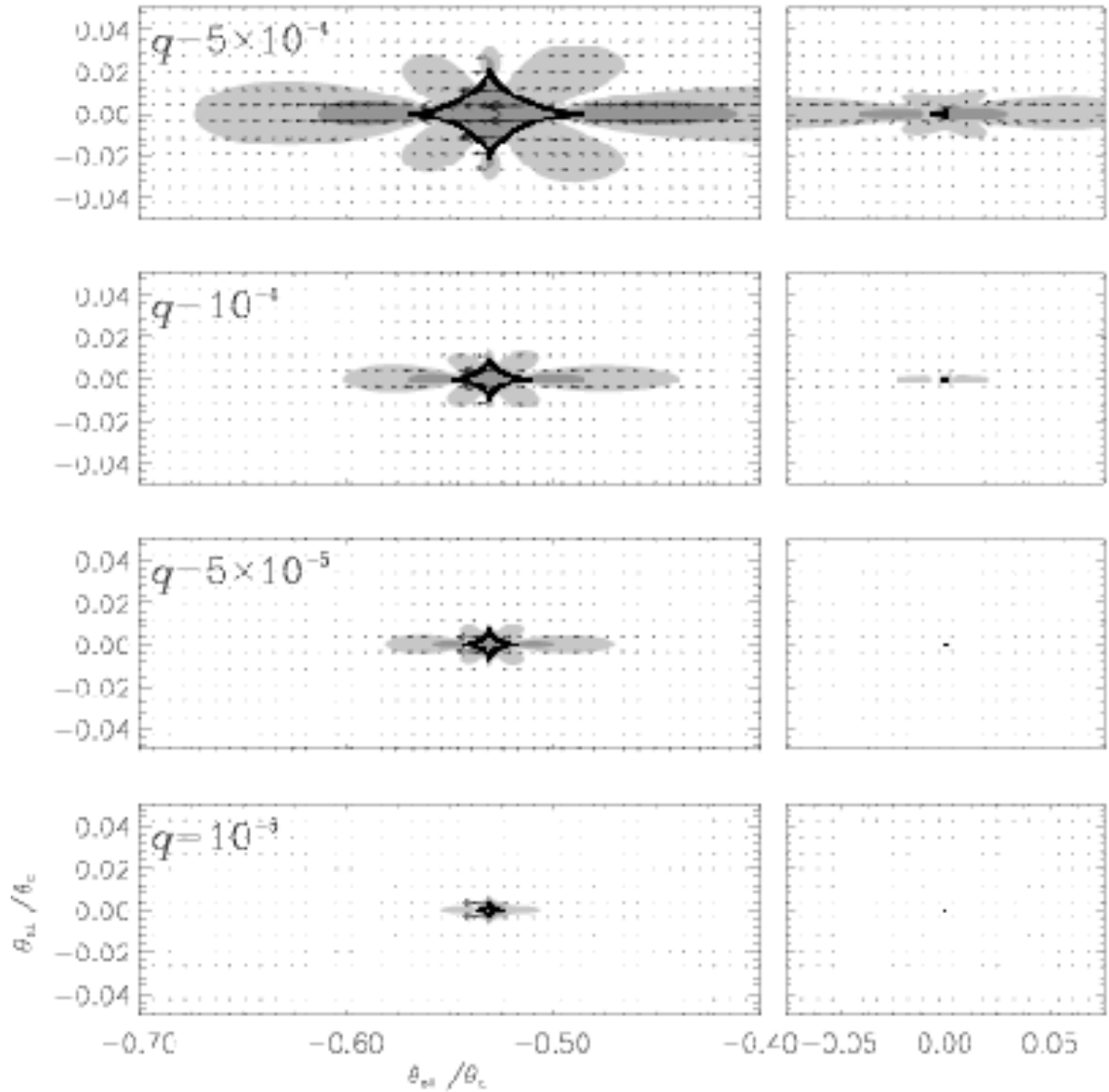}}
\caption{
Variation of the vector field map of $\Delta\deltavec$ with the
planet/primary mass ratio $q$.  The maps are similar to those presented in
Fig.\ 12, but for lens systems with smaller mass ratios.
Note that the scale of the maps are different from that of the maps
presented in Fig.\ 12.
}
\end{figure*}

\begin{figure*}
\epsfysize=17cm
\centerline{\epsfbox{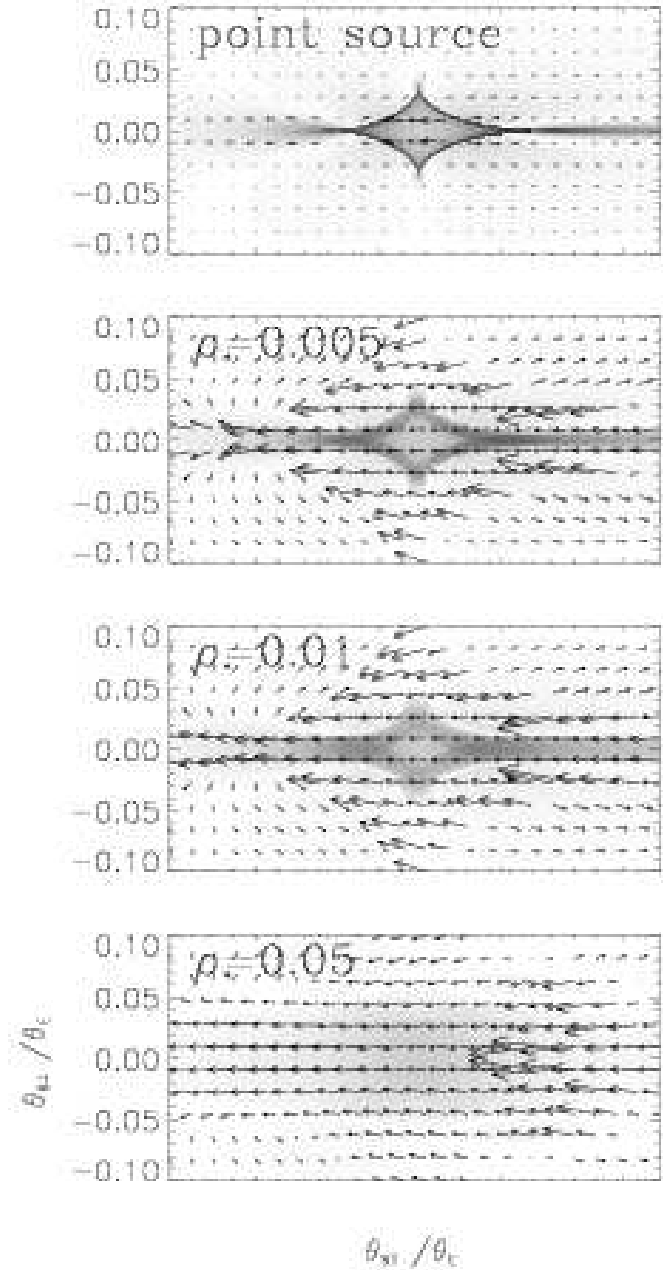}}
\caption{
Variation of the vector field map of $\Delta\deltavec$ with source size.
Unlike other maps, we do not draw caustics on these maps to better show 
the detailed structures near the caustics.  We draw 60 grey-scales in the 
range of $\Delta\delta=0.03\theta_{\rm E}$ -- $0.7\theta_{\rm E}$.
}
\end{figure*}

\end{document}